# Nuclear excitation functions from 40-200 MeV proton irradiation of terbium

Jonathan W. Engle, Stepan G. Mashnik, Lauren A. Parker, Kevin R. Jackman, Leo J. Bitteker, John L. Ullmann, Mark S. Gulley, Chandra Pillai, Kevin D. John, Eva R. Birnbaum, and Francois M. Nortier

*Los Alamos National Laboratory, Los Alamos*
*P.O. Box 1663, Los Alamos, New Mexico, 87545*

LA-UR-15-23803

e-mail address: jwengle@lanl.gov

**Abstract:** Nuclear formation cross sections are reported for 26 radionuclides, measured with 40 to 200 MeV proton irradiations of terbium foils. These data are relevant to the production of medically relevant radionuclides (e.g., $^{152}$Tb, $^{155}$Tb, $^{155}$Eu, and $^{156}$Eu) and to ongoing efforts to characterize stellar nucleosynthesis routes passing through long-lived intermediaries (e.g., $^{153}$Gd). Computational predictions from the ALICE2011, CEM03.03, Bertini, and INCL+ABLA codes are compared with newly measured data to contribute to the ongoing process of code development, and yields are calculated for selected radionuclides using measured data.

**PACS number(s):** 24.10.-i, 25.40.Sc, 87.56.bd

# I. INTRODUCTION

Despite potential to describe formation of several interesting radionuclides, few published measurements of cross sections for proton-induced reactions on terbium exist. Most such reactions, which provide access to terbium, gadolinium, and europium radioisotopes, require incident energies that exceed the capabilities of all but a few facilities whose focus is on radionuclide production. The relevant region of the Chart of the Nuclides is reproduced in Fig. 1. Terbium offers the convenience of a monoisotopic target material, requiring no isotopic enrichment to reduce the incidence of undesirable nuclear reactions. Steyn and coauthors recently described the formation of several radioisotopes of terbium, dysprosium, and gadolinium using protons below 66 MeV [1]. They were motivated by work at the Paul Scherrer Institute in Switzerland, which makes a strong case for the use of $^{152}$Tb ($t_{1/2}$ 17.5 h, 17% $\beta^+$) and $^{155}$Tb ($t_{1/2}$ 5.32 d, 32% 86.55 keV $\gamma$, 25.1% 105.318 keV $\gamma$) in positron emission tomography and single photon emission tomography imaging, respectively, and $^{149}$Tb ($t_{1/2}$ = 4.118 h, 83.3% εc, 16.7% α) and $^{161}$Tb ($t_{1/2}$ 6.89 d, 100% $\beta^-$) for radiotherapy [2]. Radioisotopes of any element that offer both diagnostic and therapeutic potential are highly sought-after because they obviate the need to consider differences in the *in vivo* kinetics of drugs radiolabeled with different elements. Reactions that form $^{149}$Tb and $^{152}$Tb from terbium targets eluded this report, as they require higher energy protons. Higher energies may enable production of other useful radionuclides besides $^{149}$Tb and $^{152}$Tb. For example, a source of $^{153}$Gd ($t_{1/2}$ 240.4 d, 100% εc) sufficiently pure of $^{151}$Gd could likely be used in measurements of its neutron capture cross section for characterization of stellar nucleosynthesis by the *S*-process. Recent publications codify this interest and estimate that the Detector for Advanced Neutron Capture Measurements (DANCE) at Los Alamos National Laboratory could make such a measurement with $10^{14-16}$ atoms $^{153}$Gd [3]. The long half-life of $^{146}$Gd ($t_{1/2}$ 48.27 d, 100% εc) suggests its use as the parent in a generator system providing a relatively pure source of positron-emitting europium. Its $^{146}$Eu daughter ($t_{1/2}$ 4.61 d, 4.6% $\beta^+$) has a larger positron branching ratio than any other radioisotope of europium that also has a half-life suitable for medical applications.

In the past, we have reported measurements of proton induced reactions on terbium at 800 MeV [4]. Additional reports consider primarily high-energy reactions (>600 MeV) [5–7], though Alexandrov and coauthors have examined formation of the alpha-emitting radionuclide $^{149g}$Tb ($t_{1/2}$ 4.118 h, 83.3% εc, 16.7% α) from a variety of elemental targets and protons with energies from 60 MeV to several GeV [8]. Following these works, a gap in published data between 66 and approximately 600 MeV remains. The Los Alamos Isotope Production Facility (IPF) and the Brookhaven Linear Isotope Producer (BLIP) routinely use proton beams of up to 200 MeV and up to 230 μA intensities to make radionuclides available through the U.S. Department of Energy's Office of Science. Work reported here attempts to fill the gap in measured data in the energy ranges used at IPF and BLIP.

As with any radionuclide production scheme, accurate nuclear excitation functions are necessary to predict yields and purities achievable given targetry and accelerator capabilities. The relative similarity in half-lives common in this region of the Chart of the

Nuclides, as well as the expected energetic overlap in excitation functions for similar nuclear reactions, creates a challenge for the physicist. Steyn and coauthors were forced to conclude that electromagnetic separation offers the only hope of obtaining radioisotopically pure $^{155}$Tb for studies of its therapeutic potential [1]. Production of other radionuclides proposed above will be subject to similar constraints – odd-mass isotopes of gadolinium will confound measurement of $^{153}$Gd's cross section, and the presence of $^{149}$Gd in a $^{146}$Gd/$^{146}$Eu generator system will contribute longer-lived $^{149}$Eu ($t_{1/2}$ 93.1 d, 100% εc) to the eluent. Only measured excitation functions provide the data needed to optimize irradiations to target radionuclides of interest.

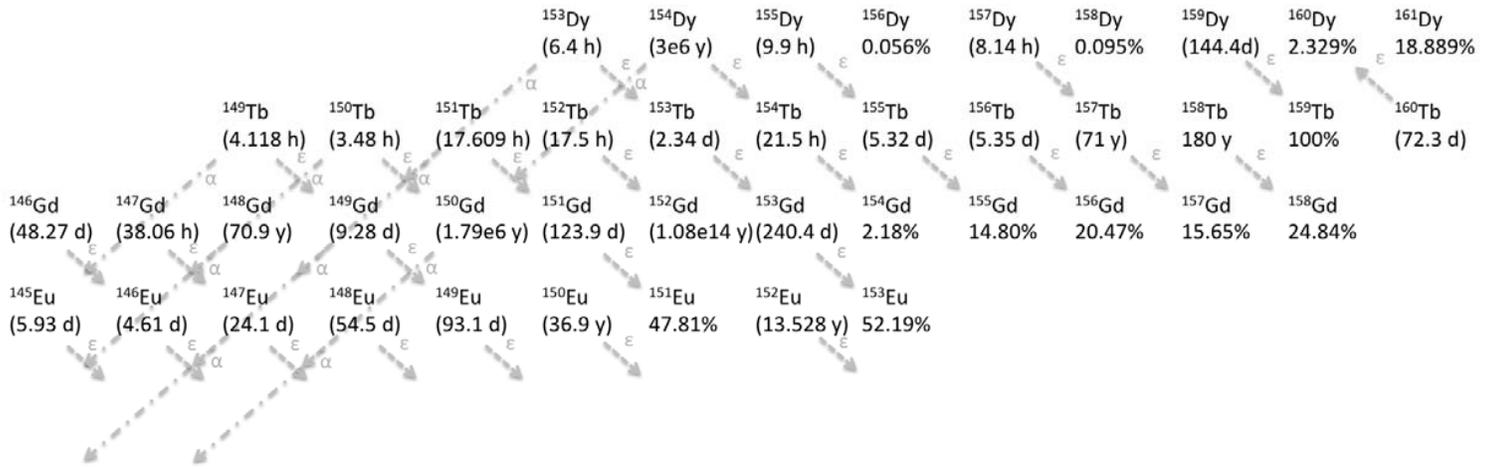

FIG. 1. Representation of the relevant region of the Chart of the Nuclides. Nuclear data are taken from [9] and additional data are summarized in Table 1.

These data are also useful for the verification and validation of nuclear physics codes that model particle transport and interaction, such as those implemented in the Monte Carlo N-Particle code, MCNP6 [10] (see, e.g., [11] and references therein) and in the ALICE code [12–15]. These codes are used in the absence of measured data for a wide variety of purposes, and improvements in their accuracy have commensurate benefit in a wide variety of fields but particularly for computational particle transport and radionuclide production.

## II. MATERIALS AND METHODS

The "stacked foil" technique is commonly employed for the measurement of charged particle nuclear excitation functions. Experimental foils are typically a few tens of mg·cm$^{-2}$ in thickness, enabling an approximately monoenergetic bombarding flux energy distribution to pass through each foil. Control of irradiation length and particle flux intensities maximizes the relative signal strength of residual radionuclides for post-irradiation spectroscopic characterization, which may be preceded by radiochemical recovery procedures. When only few or widely spread primary charged particle energies are available, thicker foils may be interspersed to degrade the primary beam energy, at the cost of increasing width in the Gaussian energy distribution of particles incident on foils towards the rear of the stack. Since nuclear formation excitation functions are generally smooth at energies above a few MeV and for targets of intermediate mass like terbium, the measurements made at these monoenergetic energies may be fitted by continuous curves, especially with semi-empirical shapes guided by theoretical models, in order to construct continuous functions that allow accurate prediction of the desired yields and purities given a set of irradiation conditions.

**A. Irradiation and gamma-ray spectrometry**

Thin terbium discs (nominally 90 mg·cm$^{-2}$) were irradiated in two experiments in the production target station of the Isotope Production Facility (IPF) at LANSCE and in the Target 2 Blue Room of the Weapons Neutron Research Facility. Both irradiations lasted approximately 1 hour and employed proton flux intensities near 100 nA, with nominal incident energies of 100 and 200 MeV, respectively. Aluminum plates were used to degrade the energy of the incident beam to allow well-spaced measurements down to approximately one half the beam's incident energy. Thin (20-50 mg·cm$^{-2}$) aluminum foils were irradiated simultaneously in order to use published values for the $^{27}$Al(p,x)$^{22}$Na reaction [16] as a monitor of integrated beam current. Stainless steel foils were also irradiated and exposed to Gafchromic film in order to confirm the beam's incidence on terbium and aluminum targets. Beam energy at each foil in the stack was calculated with a combination of MCNP6 [10] simulations, the SRIM/TRIM nuclear code [17], and an in-house developed tool to apply the formulation proposed by Anderson and Ziegler [18], as described previously [19]. Calculated proton attenuation through the stack was calculated using MCNP6 and SRIM/TRIM and compared with the measured fluences derived from aluminum monitor foils.

Following irradiation, samples were transported to the LANL Chemistry Division Countroom, where they were repeatedly assayed by non-destructive gamma-ray spectrometry for approximately 200 days. The HPGe detector used to assay the foils is a p-type aluminum windowed ORTEC GEM detector with a relative efficiency at 1333 keV of about 10% and a measured gamma peak FWHM at 1333 keV of 1.99 keV. Contributions to spectra backgrounds, detector resolution, and energy calibration (gain), were checked daily. Detector efficiency was calibrated prior to the beginning of data collection and verified after the experiment's completion. The commercial UNIX/Linux implementation of the SAMPO code originally developed by Routii and incorporated into Countroom server algorithms was used to analyze collected gamma spectra [20–23]. Gamma-ray energies and intensities listed in Table 1 were taken from the National Nuclear Data Center's (NNDC) online archives [9]. The activity at the end of bombardment (EoB) of each isotope of interest was determined by fitting of its decay curve, and cross sections were calculated using the well-known activation formula.

Uncertainties in linear regressions' fitted parameters were computed from covariance matrices as the standard deviation in the activity extrapolated to the end of bombardment. This value was combined according to the Gaussian law of error propagation with estimated contributing uncertainties from detector calibration and geometry reproducibility (5.9% combined), target foil dimensions (0.1%), and proton flux (6-8%).

Multiple photopeaks were used (up to a maximum of 4) when possible, and so additional uncertainty as the standard deviation of these complimentary measurements was combined with the uncertainties described above, again according to the Gaussian law of error propagation.

[TABLE I]

**B. MCNP6 event generators tested here**

We compare predictions of the Monte Carlo formulations of the hybrid/geometry dependent hybrid model nuclear code ALICE [12–15] and the predictions of three codes implemented as event generators in the transport code MCNP6 [10] with our newly measured data. All predictions were made prior to the measurement.

ALICE2011 is a Monte Carlo formulation based on a series of older Hybrid models of precompound decay formulated in the codes ALICE and HMS-ALICE. These codes were developed initially by Marshall Blann at Lawrence Livermore National Laboratory [13], and later in collaboration with researchers in Los Alamos, Karlsruhe, and Obninsk. It is based on HMS precompound decay [26], Weisskopf-Ewing evaporation [24], and Bohr-Wheeler [25] fission models. The latter may be run in S-wave approximation to estimate angular momentum effects on phase space, including enhancement of gamma ray de-excitation. Multiple emission cascades including photons, n, p, d, t, $^3$He, and $^4$He are considered, as well as the fission channel. Product yields may be calculated, including those of fission fragments. Single and double differential emission spectra are calculated, and the user may select to have ENDF format spectra for 1-3 n, p, and $^4$He particle emissions' output. Treatment of angular distributions uses the linear momentum conservation model of Chadwick and Oblozinsky [27]. Isomer yields are determined from the arrays of excitation vs. angular momenta for those nuclei with insufficient energy for further emission of n, p, or α [14,15]. The spins of fission fragments are assumed to result from two sources: the angular momenta of the composite nucleus that undergoes fission and an angular momentum imparted from the fission process to the fragments. These two results are coupled assuming a 2J+1 weighting among possible final spins [14,15].

The fission cross sections and probabilities are calculated using the Bohr-Wheeler model [25]. The mass and charge divisions, as well as fission fragment excitations and channel energies come from routines developed and written by Mashnik, Gudima, and collaborators for CEM03.01 [28], based on the initial work of Atchison [29], modified first by Furihata [30]. When the atomic number of an excited nucleus is less than 13, ALICE2011 uses the Fermi break-up model [31] to calculate its disintegration with a routine adopted from CEM03.01 [28]. The random number generator in ALICE2011 was also adopted from CEM03.01 [28].

ALICE2011 is intended to be relatively fast in execution and easy to use and allows the user to calculate reactions induced by both elementary particles and nuclei at incident energies up to about 250 MeV. A priority of ALICE2011 in comparison with intra-nuclear cascade (INC) type models able to describe heavy-ion induced reactions, like LAQGSM03.03 (see details in [11]) used as event generators in MCNP6 [10], is that ALICE2011 is assumed to work well for nucleus-nucleus reactions at low energies, of only several MeV/nucleon, while INC-type models are expected to be reliable for such reactions only at much higher energies, above tens of MeV/nucleon and even higher.

A brief description of the three MCNP6 event generators follows:

1) The default MCNP6 option, which for our reaction is an improved version of the Cascade-Exciton Model (CEM) of nuclear reactions as implemented in the code CEM03.03 [11,32]. CEM03.03 assumes calculated nucleon-induced reactions involve three stages: The first stage is the intranuclear cascade (INC), in which primary particles can be re-scattered and produce secondary particles several times prior to absorption by (or escape from) the nucleus. When the cascade stage of a reaction is completed, CEM03.03 uses the coalescence model to "create" high-energy d, t, $^3$He, and $^4$He particles by final-state interactions among emitted cascade nucleons. The emission of the cascade particles determines the particle–hole configuration, Z, A, and the excitation energy that is the starting point for the second, preequilibrium stage of the reaction. The subsequent relaxation of the nuclear excitation is treated with an improved version of the modified exciton model of preequilibrium decay followed by the equilibrium evaporation/fission stage (also called the compound nucleus stage), which is described with an extension of the Generalized Evaporation Models (GEM) code, GEM2, by Furihata [30]. Generally, all components may contribute to experimentally measurable particle emission spectra and affect the final residual nuclei. But if the residual nuclei after the INC have atomic numbers in the range A < 13, CEM03.03 uses the Fermi breakup model [31] to calculate their further disintegration instead of using the preequilibrium and evaporation models. Fermi breakup is faster to calculate than GEM and gives results similar to the more detailed models for lighter nuclei.

2) The Bertini IntraNuclear Cascade (INC) [33], followed by the Multistage Preequilibrium Model (MPM) [34], followed by the evaporation model as described with the EVAP code by Dresner [35], followed by or in competition with the RAL fission model [29] if the charge of the compound nucleus Z is ≥ 70, referred to herein simply as "Bertini". The Bertini default option of MCNP6 also accounts for Fermi breakup of excited nuclei when A < 18, but does not account for the coalescence of complex particles from INC nucleons.

3) The IntraNuclear Cascade model developed at the Liege (INCL) University in Belgium by Prof. Cugnon with his coauthors from CEA, Saclay, France [36] merged with the evaporation-fission model ABLA [37] developed at GSI, Darmstadt, Germany, referred to herein as "INCL+ABLA". The version of INCL + ABLA available currently in MCNP6 accounts for possible fission of compound nuclei produced in our reaction, but it does not account for preequilibrium processes, for Fermi break-up of light residual nuclei, or for coalescence of complex particles after (or during) INC.

All event generators used compute only independent cross sections; cumulative cross sections were subsequently calculated using these independent values summed separately according to the decay behavior of parent products using the Chart of the Nuclides [9].

## III. RESULTS AND DISCUSSION

**A. Cross sections**

The results of cross section measurements are presented below and describe the formation of terbium, dysprosium, gadolinium and europium radioisotopes from proton irradiations of terbium. Agreement between measured and calculated cross sections is generally good. Steyn et al. [1] is notable for its contribution of data to many of the reactions also studied in this work; agreement between these two datasets is also quite good in the energy range where there is overlap.

1. Cross sections for radioisotopes of dysprosium

Measured data for $^{153}$Dy, $^{155}$Dy, $^{157}$Dy, and $^{159}$Dy are tabulated in Tab. II and plotted in Figs. 2-5 below together with values measured by Steyn and coauthors [1] where possible. It was impossible to discern sufficient signal from the 99.66 and 659.84 keV γ-emissions of $^{153}$Dy at lower proton energies to achieve overlap with the data reported by Steyn and coauthors. Disagreement with the Steyn data is greater for the 59 MeV data point of $^{155}$Dy than for any other point reported in this work, and data for $^{157}$Dy and $^{159}$Dy are plotted on logarithmic scales in order to show the size of this discrepancy, which does not exceed 20%. Steyn and coauthors have previously described the effect of irradiation parameters and the timing of radiochemical isolation efforts on the achievable purity of $^{155}$Dy produced for SPECT applications.

[TABLE II]

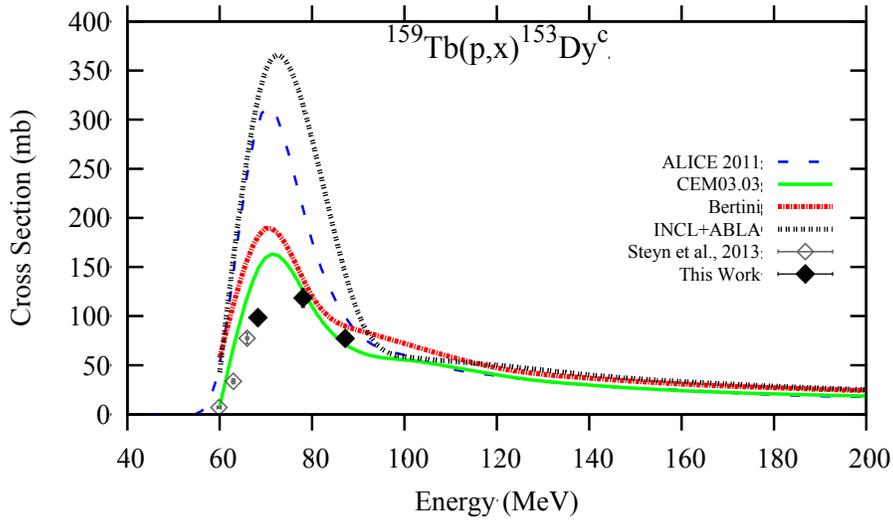

FIG. 2 (color online). Measured excitation function for the production of $^{153}$Dy by proton irradiation of Tb. Solid diamonds: this work. Open diamonds: [1]. Predictions of ALICE2011 [12,13,38], CEM03.03 [11,32], Bertini [29,33–35], and INCL+ABLA codes [36,37], the latter three implemented as event generators in MCNP6, are shown as smoothed lines for comparison.

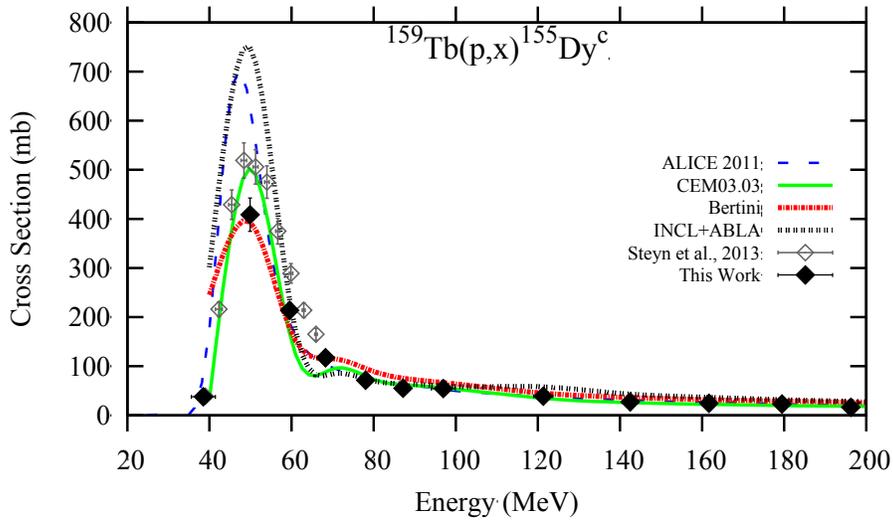

FIG. 3 (color online). The same as for Fig. 2, but for the cumulative production of $^{155}$Dy.

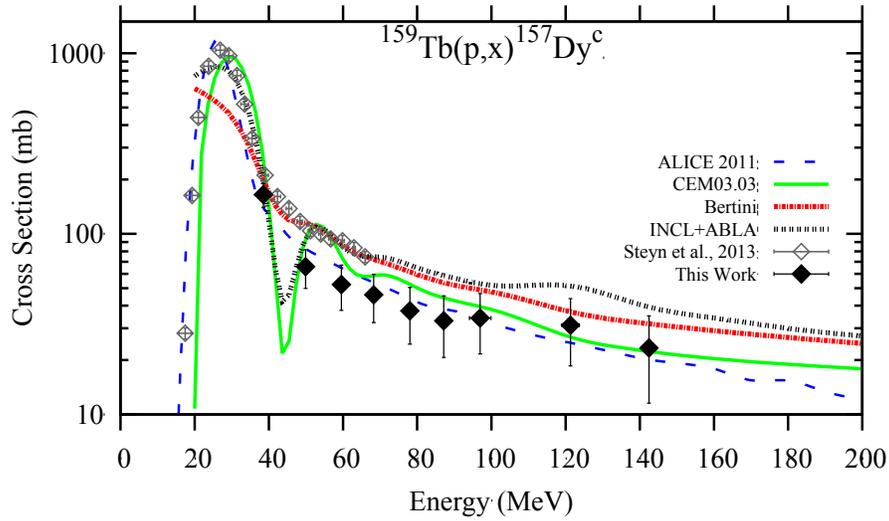

FIG. 4 (color online). The same as for Fig. 2, but for the cumulative production of $^{157}$Dy.

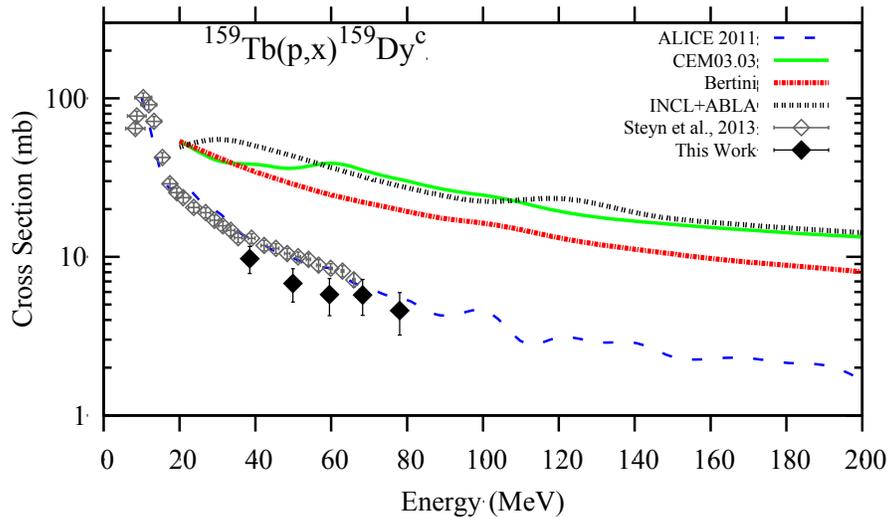

FIG. 5 (color online). The same as for Fig. 2, but for the cumulative production of $^{159}$Dy.

2. Cross sections for radioisotopes of terbium

Measured data are tabulated in Tab. III and plotted in Figs. 6-14. Where an unstable dysprosium parent exists, the majority contribution to the excitation functions for terbium isotopes reported below comes from (p,xn)-type reactions. As terbium isotope mass increases, the relative contribution of (p,pxn) reactions increases, and the predicted excitation functions exhibit the broad shape characteristic of these reactions, with a lack of the more defined peak form seen in (p,xn). No single theoretical model distinguishes itself in the prediction of cross sections for the formation of terbium radioisotopes. The shape of the measured $^{151}$Tb excitation function lacks the sharp peak predicted by both ALICE and INCL+ABLA and the trough above the peak predicted only by INCL+ABLA. All codes studied closely replicate the magnitude of the tail of the $^{152}$Tb cross section, but in many cases measured data suggest that the peak of the cross section occurs between 5 and 20 MeV higher than their predictions. In the case of the cumulative excitation function of $^{153}$Tb, the five measured data points by Steyn and coauthors are in good agreement with new data, though again the codes predict the peak of the cross section at lower energies. Disagreement between experiment and theory is most marked for the independent excitation function of $^{153}$Tb and the cumulative excitation function of $^{154g}$Tb, where the codes generally differ from measured data by as much as a factor of two. The three metastable states of this radioisotope, combined with the absence of decay contribution from $^{154}$Dy, make this a particularly challenging residual to predict. In all cases where earlier measured data are available from Steyn and coauthors ($^{153}$Tb, $^{154g}$Tb, $^{155}$Tb, and $^{156}$Tb), experimental values are in very good agreement. Unfortunately, because of its lack of characteristic gamma emissions, $^{157}$Tb could not be observed by the counting experiments conducted here.

[TABLE III]

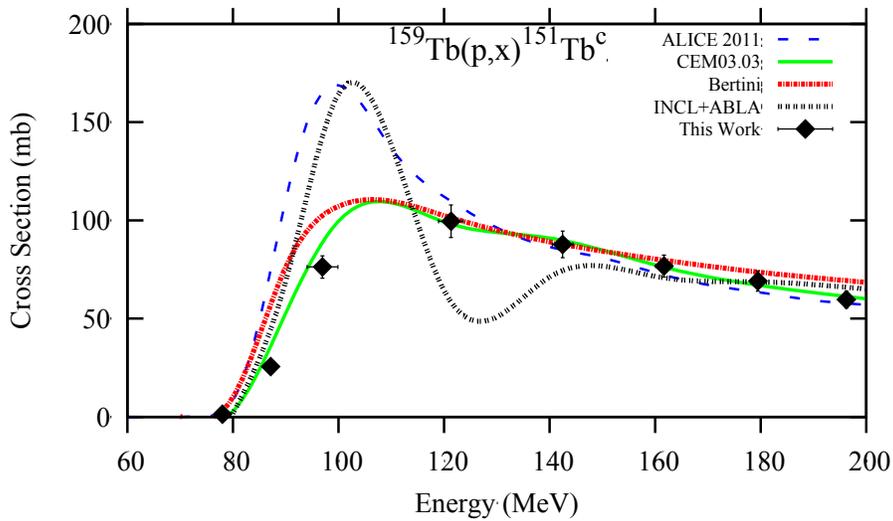

FIG. 6 (color online). The same as for Fig. 2, but for the cumulative production of $^{151}$Tb.

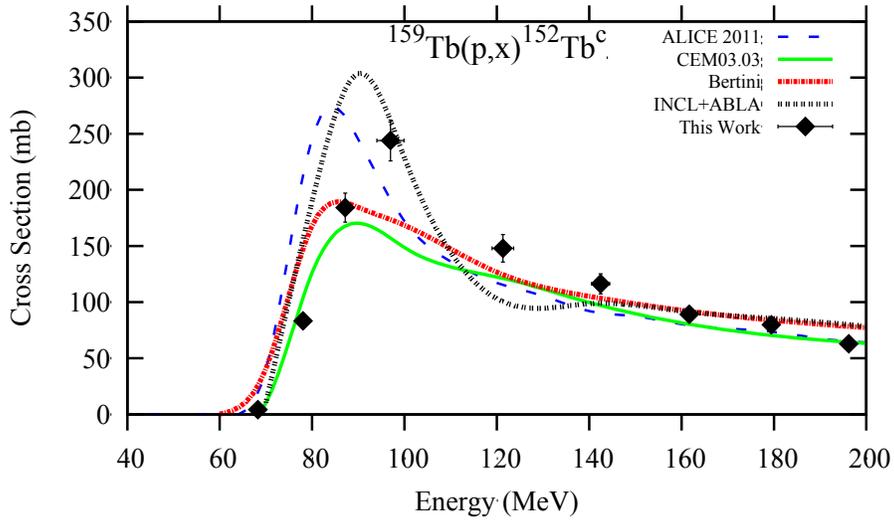

FIG. 7 (color online). The same as for Fig. 2, but for the cumulative production of $^{152}$Tb.

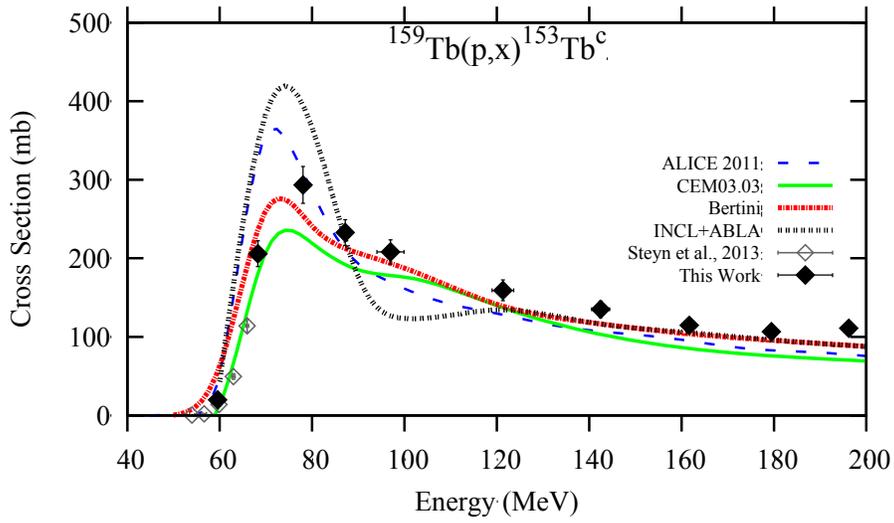

FIG. 8 (color online). The same as for Fig. 2, but for the cumulative production of $^{153}$Tb.

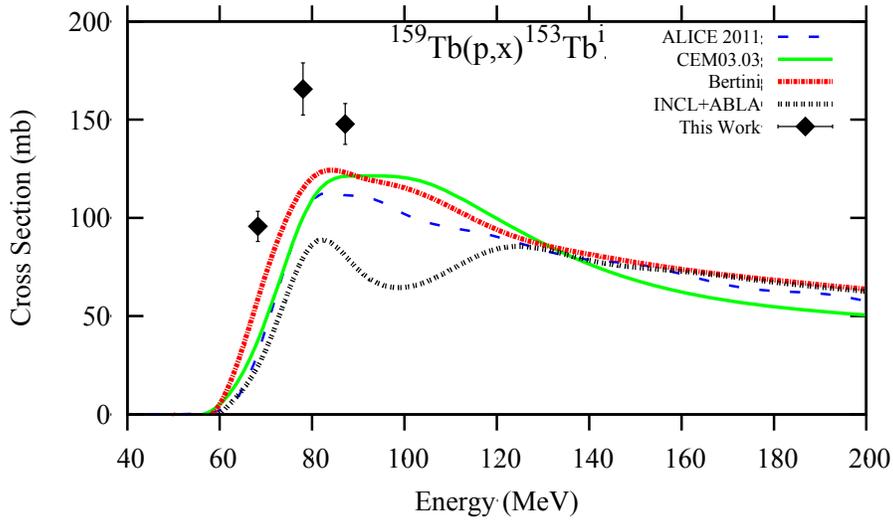

FIG. 9 (color online). The same as for Fig. 2, but for the independent production of $^{153}$Tb.

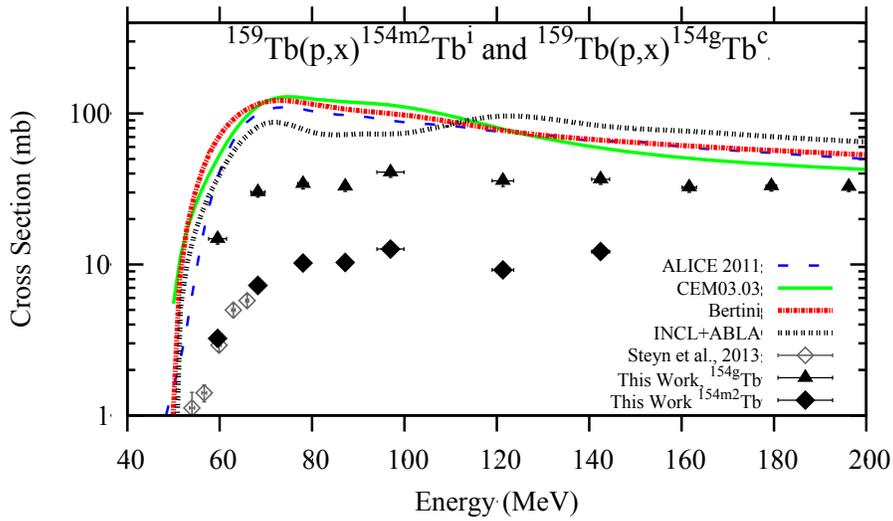

FIG. 10 (color online). The same as for Fig. 2, but for the cumulative production of $^{154m2}$Tb and $^{154g}$Tb. All codes predict only cumulative formation of the ground state.

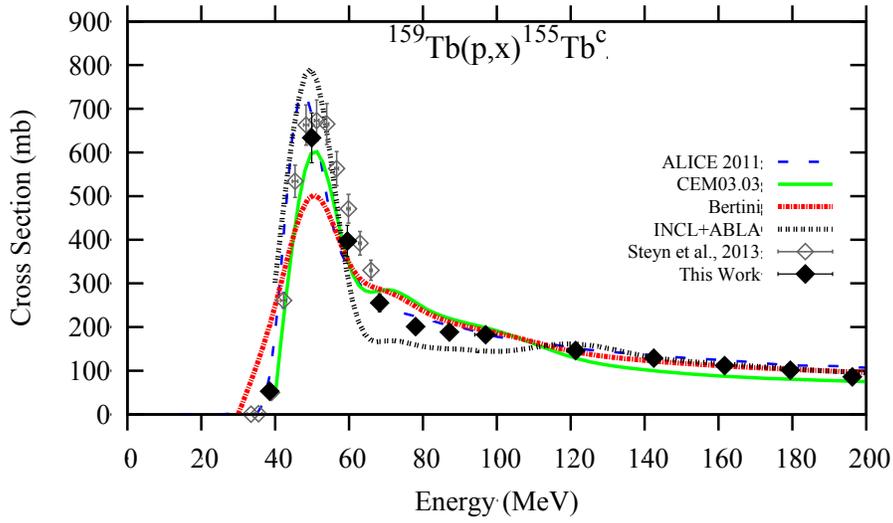

FIG. 11 (color online). The same as for Fig. 2, but for the cumulative production of $^{155}$Tb.

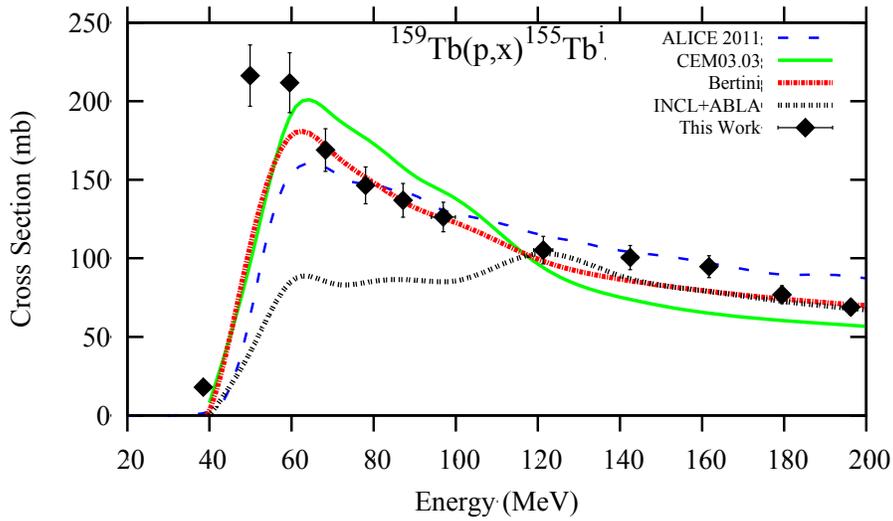

FIG. 12 (color online). The same as for Fig. 2, but for the independent production of $^{155}$Tb.

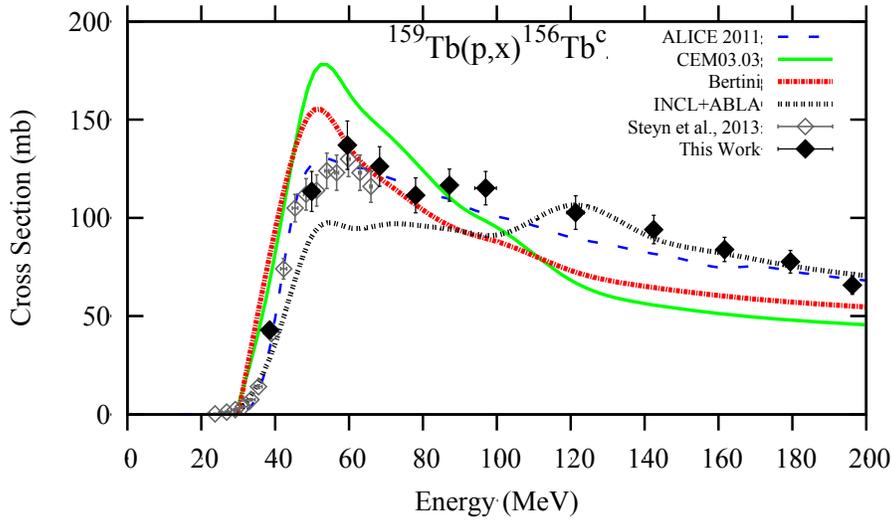

FIG. 13 (color online). The same as for Fig. 2, but for the cumulative production of $^{156}$Tb.

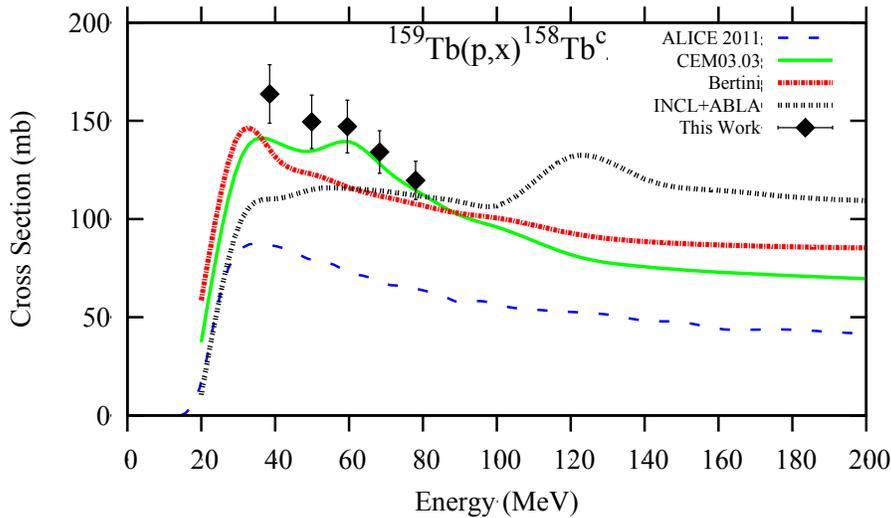

FIG. 14 (color online). The same as for Fig. 2, but for the cumulative production of $^{158}$Tb.

3. Cross sections for radioisotopes of gadolinium

Measured data for gadolinium radioisotopes are tabulated in Tab. IV and plotted along with theoretical predictions in Figs. 15-21. The observed formation of all radioisotopes of gadolinium receives contributions indirectly form (p,xn) and directly from (p,αxn) reactions, and these mechanisms dominate the distinct features in the predicted excitation functions shown below. When compared with measured data, ALICE and INCL+ABLA regularly over-estimate the contribution of (p,αxn) channels, which are generally the lowest energy features in the plotted functions. However, all codes examined agree acceptably with measured data in estimating the magnitude and shape of broader (p,xn) contribution features at higher energies. Calculation of cumulative excitation functions for $^{146}$Gd and $^{147}$Gd is additionally complicated by the significant alpha branching of their respective parents $^{150}$Dy (36.0% α) and $^{151}$Dy (5.60% α). The cumulative excitation functions of $^{151}$Gd and $^{153}$Gd were also measured by Steyn and coauthors [1], and agreement with here-measured data is acceptable but speculative, as only the 60 MeV data point provides energy overlap for comparison.

[ TABLE IV ]

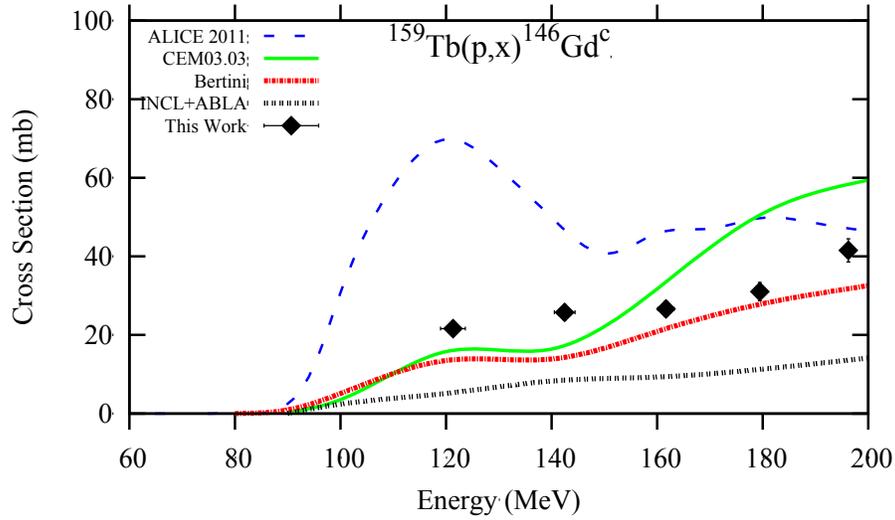

FIG. 15 (color online). The same as for Fig. 2, but for the cumulative production of $^{146}$Gd.

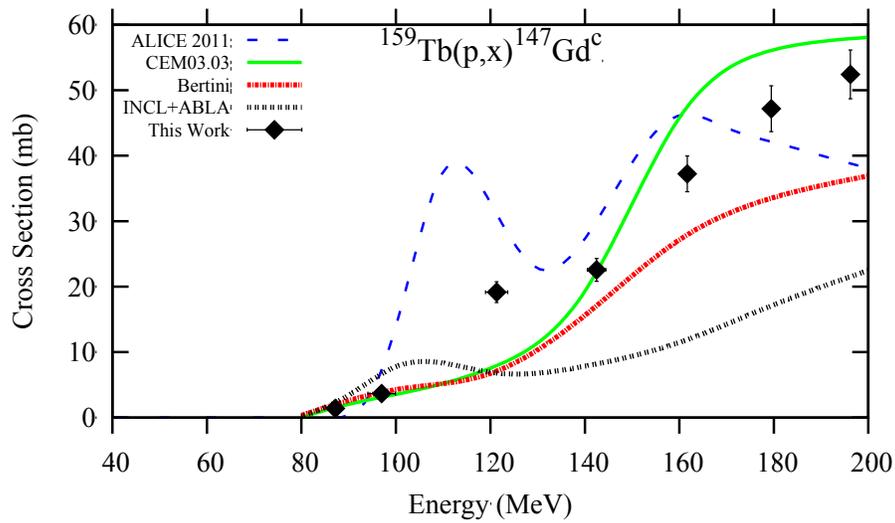

FIG. 16 (color online). The same as for Fig. 2, but for the cumulative production of $^{147}$Gd.

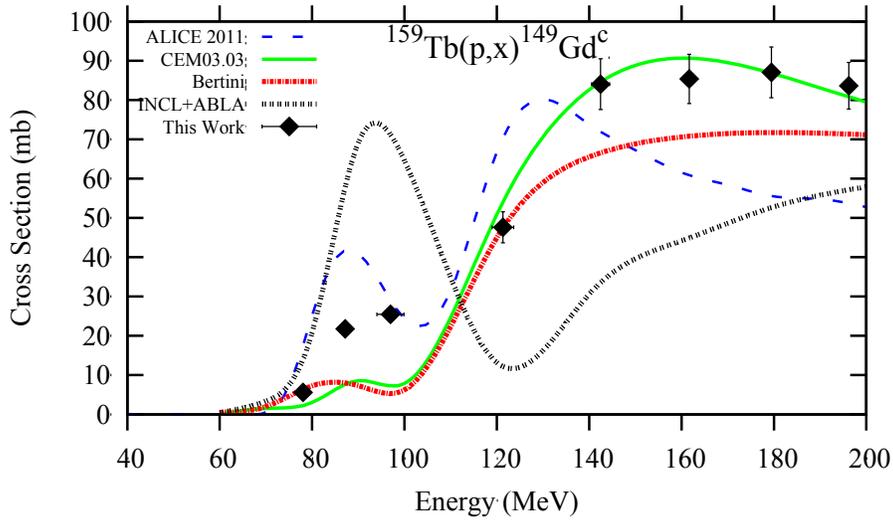

FIG. 17 (color online). The same as for Fig. 2, but for the cumulative production of $^{149}$Gd.

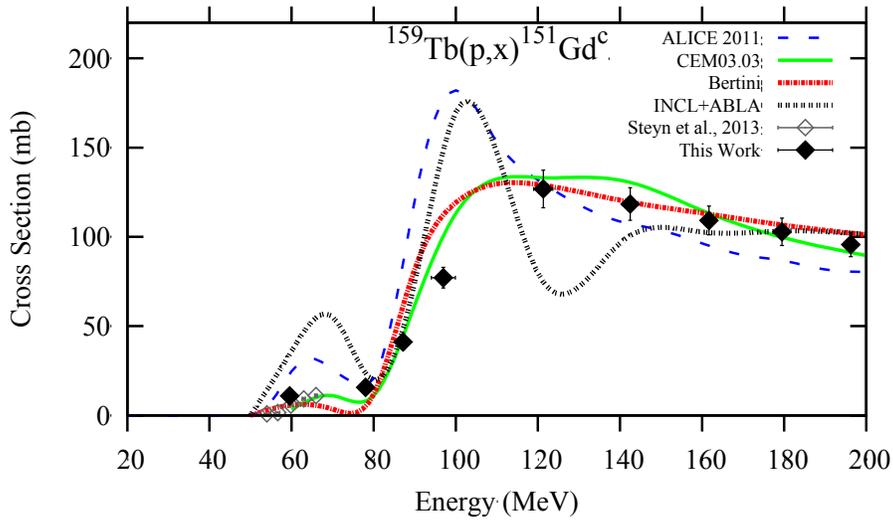

FIG. 18 (color online). The same as for Fig. 2, but for the cumulative production of $^{151}$Gd.

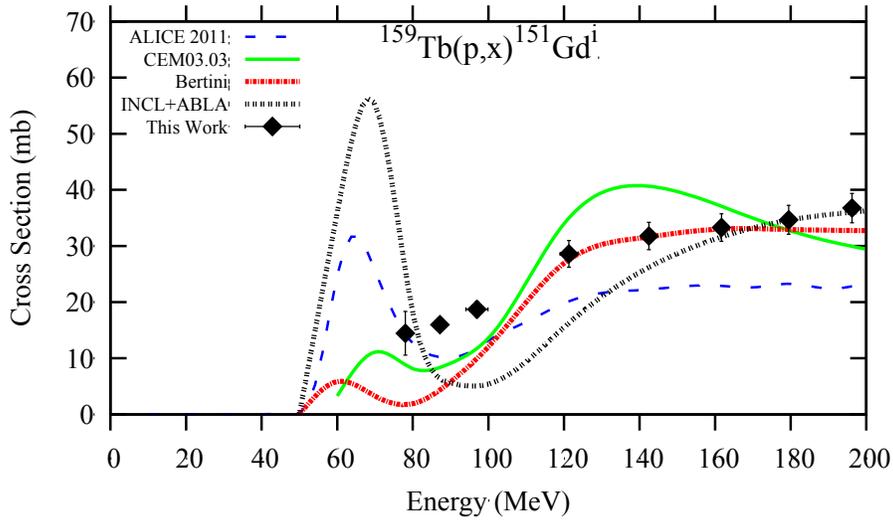

FIG. 19 (color online). The same as for Fig. 2, but for the independent production of $^{151}$Gd.

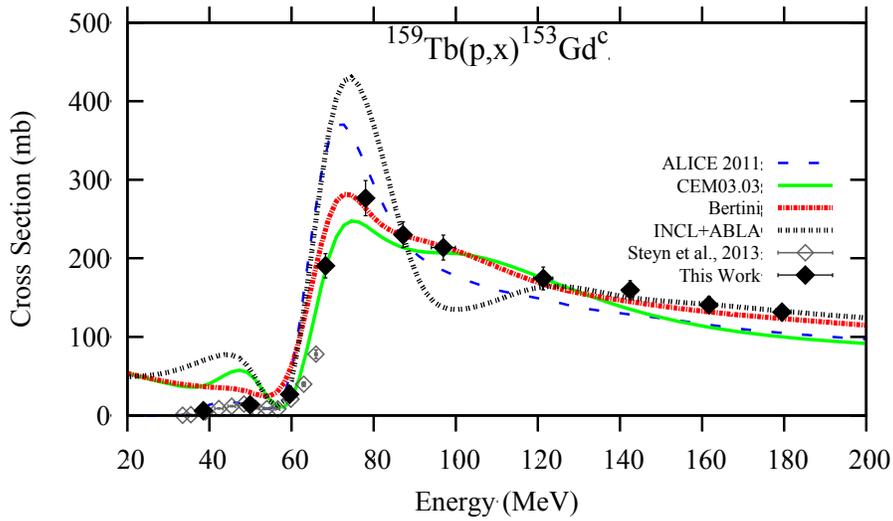

FIG. 20 (color online). The same as for Fig. 2, but for the cumulative production of $^{153}$Gd.

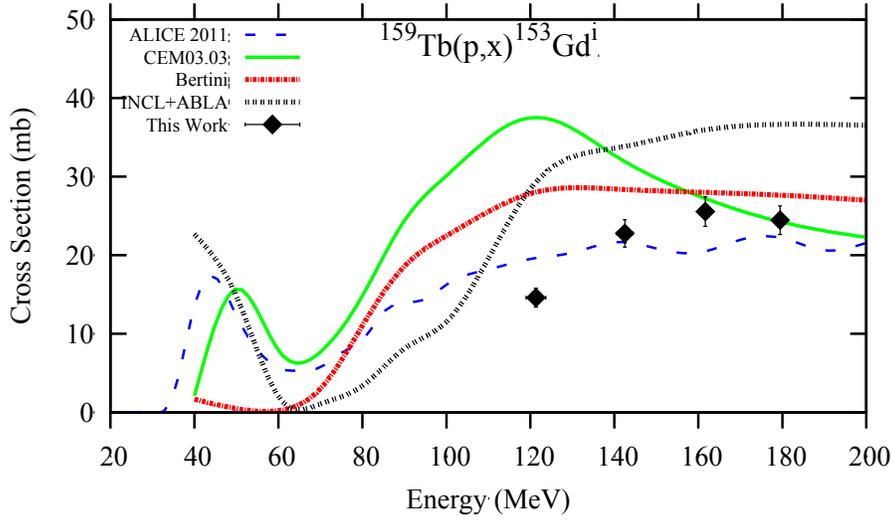

FIG. 21 (color online). The same as for Fig. 2, but for the independent production of $^{153}$Gd.

4. Cross sections for radioisotopes of europium

Measured data for europium radioisotopes are summarized in Tab. V and plotted in Figs. 22-27. Formation cross sections of these radionuclides are generally dominated by contributions from the formation and decay of their parents, discussed previously. This is illustrated in two cases, $^{146}$Gd in the case of $^{146}$Eu and $^{149}$Gd in the case of $^{149}$Eu, where the long half life of the gadolinium parent enables quantification of the europium daughter's independent formation prior to ingrowth from the parent's decay. In both instances the magnitude of the independent excitation function is approximately a factor of three smaller than that of the cumulative excitation function. The discrepancy between theoretical code predictions of distinctive shape and measured data for (p,αxn) contributions described for gadolinium radioisotopes propagates to daughter europiums (see Figs. 24, 26, and 27).

[TABLE V]

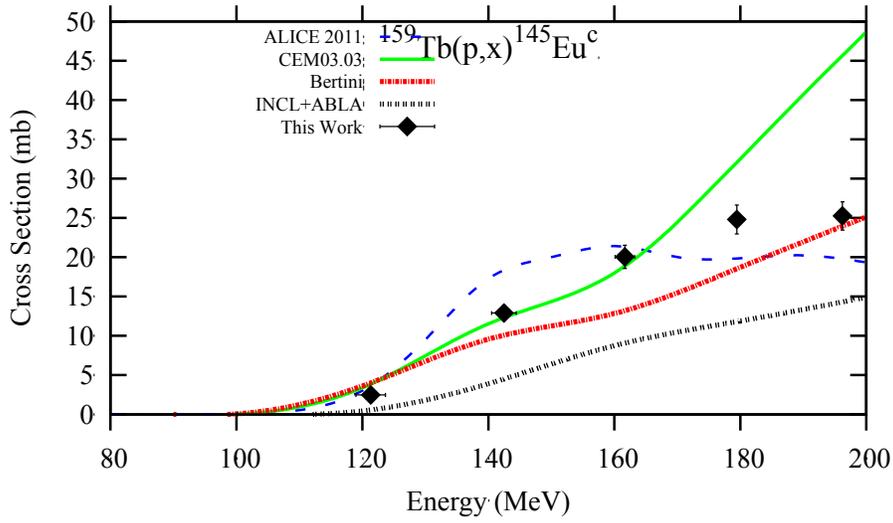

FIG. 22 (color online). The same as for Fig. 2, but for the cumulative production of $^{145}$Eu.

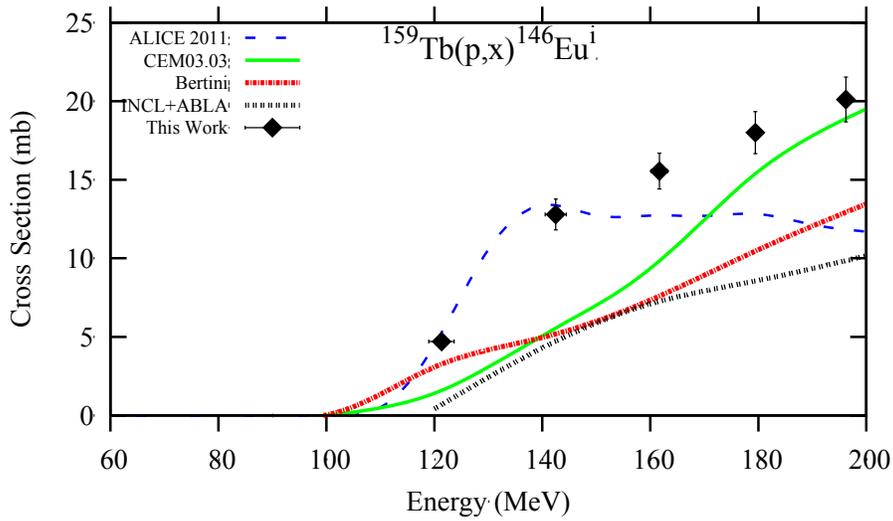

FIG. 23 (color online). The same as for Fig. 2, but for the independent production of $^{146}$Eu.

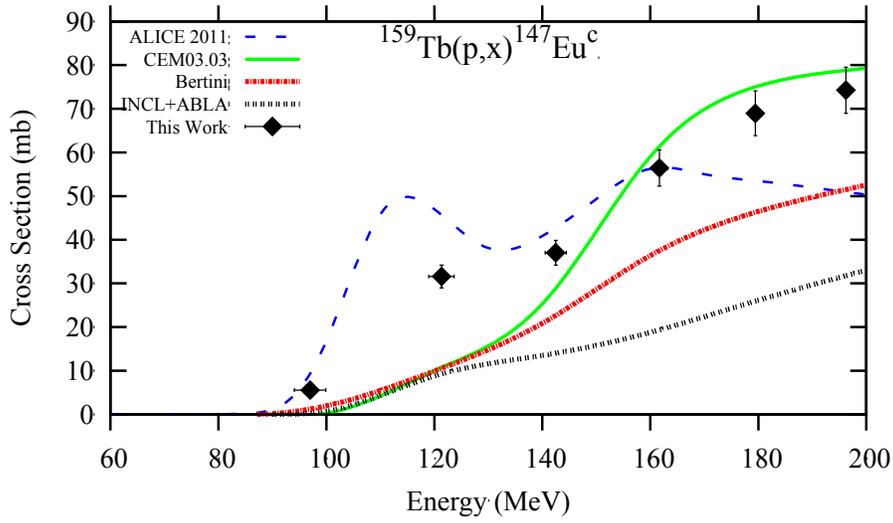

FIG. 24 (color online). The same as for Fig. 2, but for the cumulative production of $^{147}$Eu.

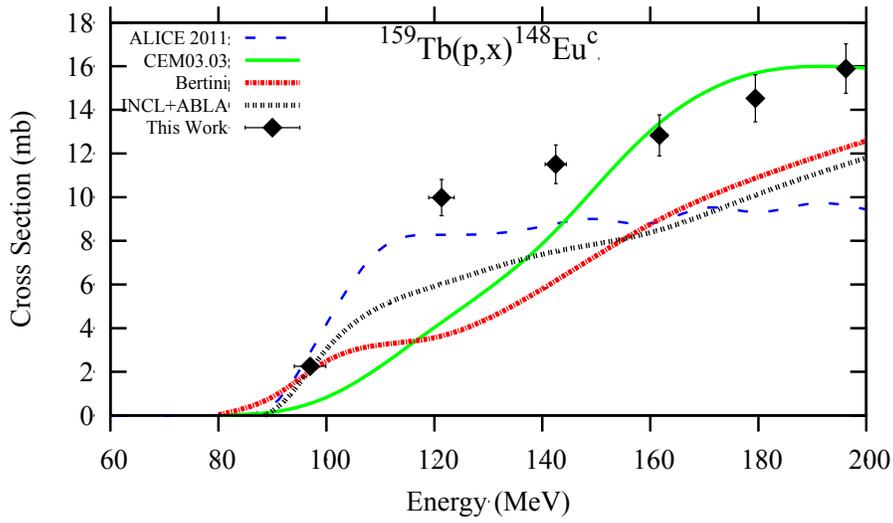

FIG. 25 (color online). The same as for Fig. 2, but for the cumulative production of $^{148}$Eu.

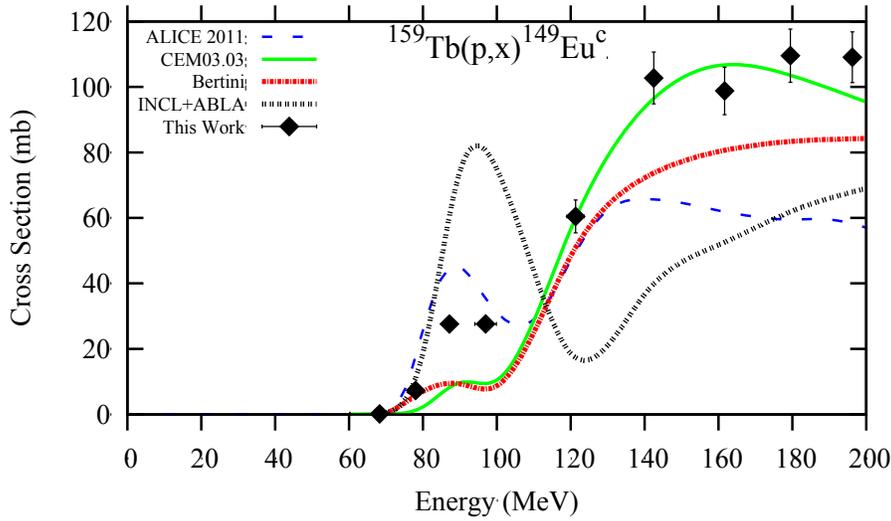

FIG. 26 (color online). The same as for Fig. 2, but for the cumulative production of $^{149}$Eu.

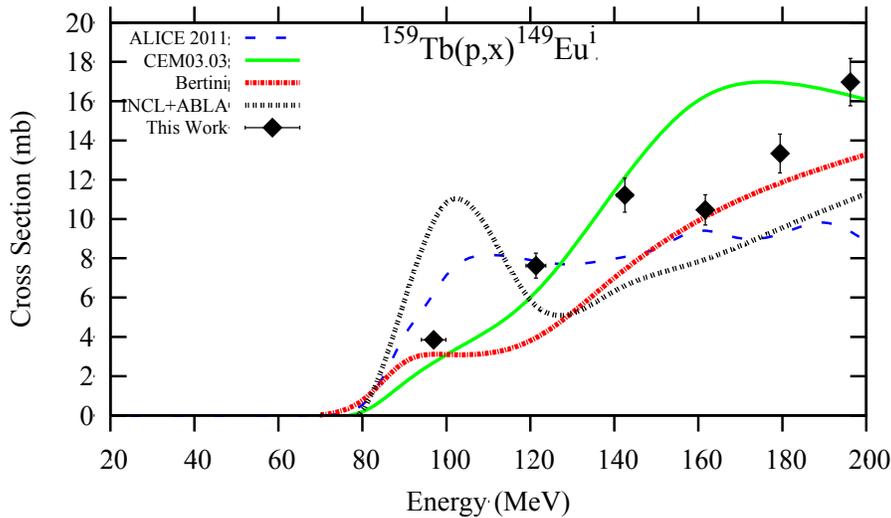

FIG. 27 (color online). The same as for Fig. 2, but for the independent production of $^{149}$Eu.

## B. Predicted yields and radiochemical purity of isotopes of interest

Thick target, or integral, yields of dysprosium radionuclides have been calculated previously by Steyn and coauthors [1] in more relevant energy ranges not measured here. Yields of terbium and gadolinium radioisotopes are plotted in Figs. 28 and 29 below. No energy range exists which might afford radioisotopically pure production of positron-emitting $^{152}$Tb, as the half lives of $^{151}$Tb and $^{153}$Tb are relatively similar, the peaks of their measured excitation functions differ by less than 20 MeV, and the functions' magnitudes are also similar.

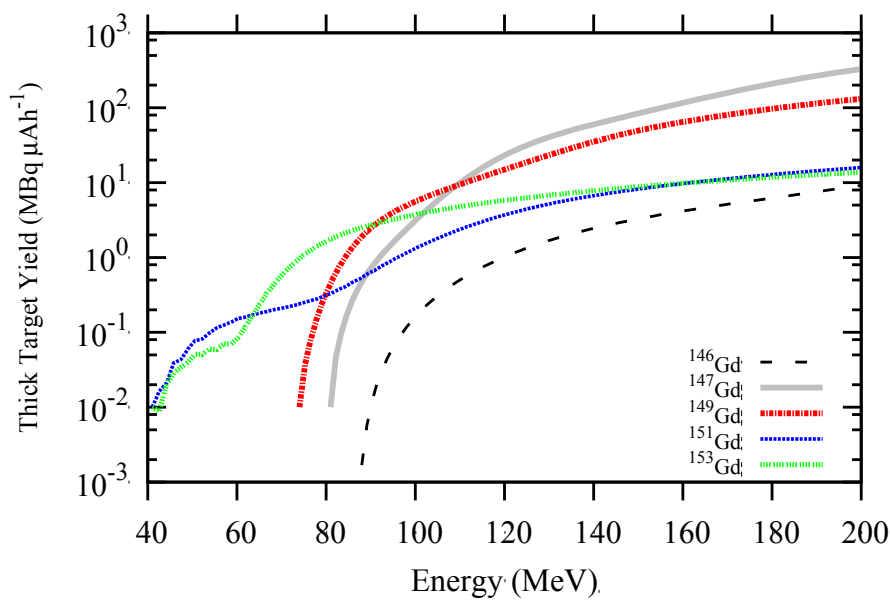

FIG. 28 (color online). Calculated here thick (integral) target yields for all measured radioisotopes of gadolinium.

Measured data suggest more promising prospects for the production of $^{153}$Gd and $^{146}$Gd/$^{146}$Eu. In the past, $^{153}$Gd has been produced for use in photon line sources, but effective commercial production has largely relied on neutron activation of gadolinium targets [39], though processes which use targets of separated $^{151}$Eu are in active development [40]. Older reactor processes which activated europium targets of natural isotopic abundance produced undesirable quantities of $^{155}$Gd from successive neutron capture reactions on the $^{153}$Eu (52.19% n.a.) present in the target [41]. A method for production of $^{153}$Gd free from the presence of odd-mass gadolinium isotopic contamination would be necessary for useful neutron capture cross section measurements on this radionuclide. Thin target yields of $^{151}$Gd, $^{153}$Gd, and $^{155}$Tb (included as the contributor of stable $^{155}$Gd), calculated from cumulative excitation functions and normalized to a unit of target thickness or energy, are plotted in Fig. 30. This data suggests use of proton energies between approximately 60 and 80 MeV where production of $^{151}$Gd may be constrained relative to $^{153}$Gd. Calculated instantaneous yields for a target with this 20 MeV thickness, which is achievable at IPF and BLIP, are 2.6 and 40.9 µCi·µAh$^{-1}$ (1.4e12 and 4.1e13 atoms µAh$^{-1}$), respectively. For a fairly routine irradiation using 240 µA proton intensity for 10 days, 10$^{18}$ atoms of $^{153}$Gd could be produced with less than 5% atomic $^{151}$Gd contamination. $^{155}$Gd contamination from formation and decay of 5.32-day $^{155}$Tb is unavoidable, but can be addressed by minimizing radiochemical processing time, maximizing the quantity of the isobar that is removed before decay to $^{155}$Gd.

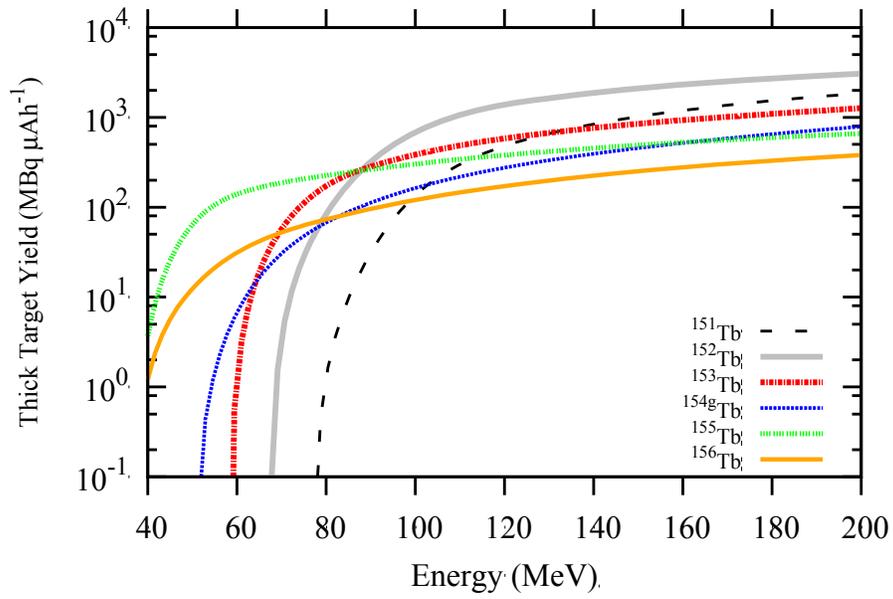

FIG. 29 (color online). Calculated here thick (integral) target yields for all measured radioisotopes of terbium.

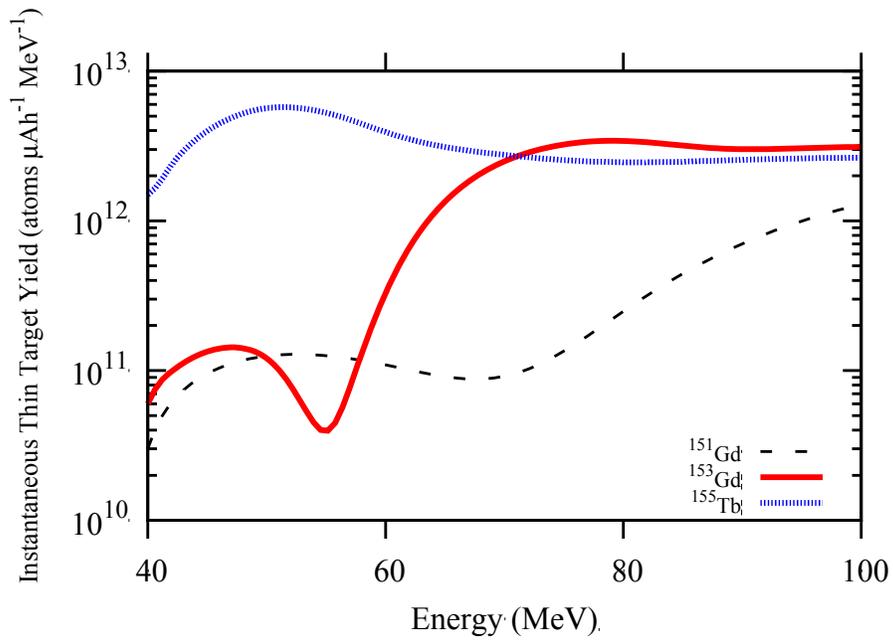

FIG. 30 (color online). Instantaneous thin target yields of odd-mass radionuclides relevant to the production of $^{153}$Gd, emphasizing the relevant energy region between 60 and 80 MeV.

A system capable of generating high yields of radioisotopically pure, positron-emitting $^{146}$Eu could be useful in the development of new diagnostic radiotracers for PET. After a suitable waiting period following irradiation to allow shorter-lived radionuclides to decay, an initial radiochemical separation of gadolinium and europium would isolate $^{146}$Gd and $^{149}$Gd. The energy region between 120 and 90 MeV is expected to produce approximately 27 and 350 µCi·µAh$^{-1}$ of $^{146}$Gd and $^{149}$Gd, respectively. After two half lives of the desired $^{146}$Gd, the product mixture will contain less than 3% $^{149}$Gd. With a standard 240 µA, 10 day proton irradiation at BLIP, $10^{2-3}$ MBq of $^{146}$Gd would be available for generation of $^{146}$Eu, though this production scheme cannot be accessed using the incident energy available at IPF.

## IV. CONCLUSIONS

Nuclear formation excitation functions for radioisotopes of terbium, dysprosium, gadolinium, and europium formed by 40 to 200 MeV irradiation of terbium have been measured and reported. In an effort to assist with the ongoing process of code development, the *a priori* predictions of the ALICE2011, CEM03.03, Bertini, and INCL+ABLA codes are compared with newly measured data. Medium-energy proton irradiations of terbium show promise for the production of pure sources of $^{153}$Gd and for the production of $^{146}$Gd/$^{146}$Eu generator systems of positron-emitting europium radioisotopes.

## V. ACKNOWLEDGEMENTS

We are grateful for technical assistance from LANL C-NR, C-IIAC, AOT-OPS, and LANSCE-NS groups' staff. This study was carried out under the auspices of the National Nuclear Security Administration of the U.S. Department of Energy at Los Alamos National Laboratory under Contract No. DE-AC52-06NA253996 with partial funding by the US DOE Office of Science via an funding from the Isotope Development and Production for Research and Applications subprogram in the Office of Nuclear Physics. JWE is grateful for fellowship support from the Los Alamos National Laboratory LDRD program.

TABLE 1. Nuclear data used for quantification of radionuclides, taken from [9].

| Nuclide | Half-life | Decay Mode | γ-rays (keV) | Intensity (%) |
|---|---|---|---|---|
| $^{153}$Dy | 6.4 h | ε + β$^+$: 99.99% | 99.66 | 10.51 |
| | | α: 9.4e$^-$3% | 659.84 | 1.1 |
| $^{155}$Dy | 9.9 h | ε + β$^+$: 100% | 184.56 | 3.37 |
| | | | 226.92 | 68.4 |
| | | | 271.06 | 1.21 |
| | | | 498.62 | 1.75 |
| $^{157}$Dy | 8.14 h | ε: 100% | 326.34 | 93 |
| $^{159}$Dy | 144.4 d | ε: 100% | 58 | 2.27 |
| $^{151}$Tb | 17.609 h | ε + β$^+$: 99.99% | 616.56 | 10.4 |
| | | α: 9.5e$^-$3% | 251.86 | 26.3 |
| | | | 287.36 | 28.3 |
| | | | 931.23 | 7.7 |
| $^{152}$Tb | 17.5 h | ε + β$^+$: 100% | 344.28 | 65 |
| | | α: <7e$^-$7% | 271.08 | 8.6 |
| | | | 586.29 | 9.4 |
| | | | 1299.11 | 2.15 |
| $^{153}$Tb | 2.34 d | ε: 100% | 249.55 | 2.33 |
| | | | 212 | 31 |
| | | | 170.42 | 6.3 |
| | | | 141.91 | 1.07 |
| $^{154m2}$Tb | 23 h | ε + β$^+$: 78.2% | 346.64 | 69 |
| | | IT: 21.8% | | |
| $^{154g}$Tb | 21.5 h | ε: 100% | 1996.61 | 7.5 |
| | | | 1291.33 | 6.9 |
| | | | 1274.44 | 10.5 |
| | | | 1123.09 | 5.7 |
| $^{155}$Tb | 5.32 d | ε: 100% | 180.08 | 7.5 |
| | | | 105.32 | 25.1 |
| | | | 262.27 | 5.3 |
| | | | 367.36 | 2.31 |
| $^{156}$Tb | 5.35 d | ε: 100% | 780.08 | 2.33 |
| | | | 1065.1 | 10.8 |
| $^{158}$Tb | 180 y | ε: 83.4% | 944.19 | 43.9 |
| | | β$^-$: 16.6% | 181.94 | 9.9 |
| $^{146}$Gd | 48.27 d | ε + β$^+$: 100% | 115 | 44 |
| | | | 154.57 | 45.12 |
| $^{147}$Gd | 38.06 h | ε + β+: 100% | 229.32 | 58 |
| | | | 396 | 31.4 |
| | | | 370 | 15.7 |

| Nuclide | Half-life | Decay mode | Energy (keV) | Intensity (%) |
|---|---|---|---|---|
| | | | 929.01 | 18.4 |
| $^{149}$Gd | 9.28 d | ε: 100% | 149.74 | 48 |
| | | α: 4.3e$^{-}$4% | 298.63 | 28.6 |
| | | | 346.65 | 23.9 |
| | | | 788.88 | 7.3 |
| $^{151}$Gd | 123.9 d | ε: 100% | 243.29 | 5.6 |
| | | α: ≈8.0e$^{-}$7% | 307.5 | 1.04 |
| $^{153}$Gd | 240.4 d | ε: 100% | 97.43 | 29 |
| | | | 103.18 | 21.1 |
| $^{145}$Eu | 5.93 d | ε + β$^{+}$: 100% | 893.73 | 66 |
| | | | 1658.53 | 14.9 |
| | | | 1997 | 7.2 |
| | | | 653.61 | 15 |
| $^{146}$Eu | 4.61 d | ε + β$^{+}$: 100% | 1297.03 | 5.39 |
| | | | 633 | 80.9 |
| | | | 747.16 | 99 |
| | | | 1533.71 | 6.08 |
| $^{147}$Eu | 24.1 d | ε + β$^{+}$: 100% | 121.22 | 21.2 |
| | | α: 2.2e$^{-}$3% | 197.3 | 24.4 |
| | | | 677.52 | 9 |
| | | | 601.45 | 5.42 |
| $^{148}$Eu | 54.5 d | ε + β$^{+}$: 100% | 414.00 | 20.4 |
| | | α: 9.4e$^{-}$7% | 550.28 | 99 |
| | | | 553.20 | 17.9 |
| | | | 629.99 | 71.9 |
| $^{149}$Eu | 93.1 d | ε + β$^{+}$: 100% | 327.53 | 4.03 |
| | | | 277.09 | 3.56 |

TABLE II. Measured cross sections for the production of Dy radioisotopes by proton irradiation of terbium.

| Proton Energy (MeV) | ΔE (MeV) | $^{153}$Dy$^c$ | Δσ | $^{155}$Dy$^c$ | Δσ | $^{157}$Dy$^c$ | Δσ | $^{159}$Dy$^c$ | Δσ |
|---|---|---|---|---|---|---|---|---|---|
| 38.5 | 2.9 | - | - | 37.7 | 2.8 | 164.8 | 24.8 | 9.7 | 1.9 |
| 49.9 | 1.3 | - | - | 408.6 | 34.0 | 65.8 | 15.9 | 6.8 | 1.6 |
| 59.6 | 1.1 | - | - | 213.8 | 16.4 | 52.3 | 14.7 | 5.8 | 1.5 |
| 68.3 | 1.5 | 98.5 | 7.9 | 117.3 | 8.6 | 45.9 | 13.7 | 5.7 | 1.5 |
| 78.0 | 0.7 | 118.5 | 9.5 | 71.4 | 5.3 | 37.4 | 13.0 | 4.6 | 1.4 |
| 87.2 | 0.4 | 77.3 | 5.4 | 55.1 | 3.9 | 33.0 | 12.3 | - | - |
| 97.0 | 2.9 | - | - | 54.4 | 4.0 | 34.1 | 12.5 | - | - |
| 121.3 | 2.3 | - | - | 39.0 | 3.2 | 31.2 | 12.6 | - | - |
| 142.5 | 2.0 | - | - | 27.3 | 2.1 | 23.3 | 11.8 | - | - |
| 161.7 | 1.5 | - | - | 24.5 | 1.8 | - | - | - | - |
| 179.5 | 1.1 | - | - | 23.0 | 1.7 | - | - | - | - |
| 196.2 | 0.4 | - | - | 16.2 | 1.2 | - | - | - | - |

* Superscripts ($^c$,$^i$) refer to the type of cross section measured: c = cumulative, accounting for the decay of parent radionuclides to the quantified daughter; i = independent, considering direct formation only.

TABLE III. Measured cross sections for the production of Tb radioisotopes by proton irradiation of terbium.

| Proton Energy (MeV) | ΔE (MeV) | $^{151}$Tb$^c$ | Δσ | $^{152}$Tb$^c$ | Δσ | $^{153}$Tb$^c$ | Δσ | $^{153}$Tb$^i$ | Δσ | $^{154g}$Tb$^c$ | Δσ | $^{154m2}$Tb$^c$ | Δσ | $^{155}$Tb$^c$ | Δσ | $^{155}$Tb$^i$ | Δσ | $^{156}$Tb$^c$ | Δσ | $^{158}$Tb$^c$ | Δσ |
|---|---|---|---|---|---|---|---|---|---|---|---|---|---|---|---|---|---|---|---|---|---|
| | | Cross Section (mb)* | | | | | | | | | | | | | | | | | | | |
| 38.5 | 2.9 | - | - | - | - | - | - | - | - | - | - | - | - | 53.1 | 4.8 | 18.0 | 1.6 | 42.9 | 3.9 | 163.7 | 14.9 |
| 49.9 | 1.3 | - | - | - | - | - | - | - | - | - | - | - | - | 633.6 | 57.0 | 216.3 | 19.5 | 113.5 | 10.2 | 149.5 | 13.6 |
| 59.6 | 1.1 | - | - | - | - | 19.8 | 1.8 | - | - | 14.8 | 1.1 | 3.2 | 0.3 | 397.0 | 35.8 | 211.7 | 19.1 | 137.0 | 12.3 | 147.1 | 13.4 |
| 68.3 | 1.5 | - | - | 4.2 | 0.3 | 205.8 | 16.5 | 95.7 | 7.7 | 30.2 | 2.2 | 7.3 | 0.6 | 255.5 | 20.5 | 168.9 | 13.6 | 126.2 | 10.1 | 134.1 | 10.9 |
| 78.0 | 0.7 | 1.3 | 0.1 | 83.3 | 6.7 | 293.3 | 23.5 | 165.6 | 13.3 | 34.2 | 2.5 | 10.2 | 0.8 | 201.1 | 16.1 | 146.4 | 11.7 | 111.5 | 8.9 | 119.7 | 9.7 |
| 87.2 | 0.4 | 25.6 | 1.8 | 184.1 | 12.9 | 232.8 | 16.3 | 147.8 | 10.4 | 32.8 | 2.3 | 10.3 | 0.7 | 188.4 | 13.2 | 136.9 | 10.7 | 116.6 | 8.2 | - | - |
| 97.0 | 2.9 | 76.3 | 5.7 | 244.0 | 18.1 | 208.1 | 15.5 | - | - | 40.8 | 3.0 | 12.6 | 0.9 | 182.5 | 13.6 | 126.3 | 9.4 | 115.1 | 8.6 | - | - |
| 121.3 | 2.3 | 99.6 | 8.3 | 147.9 | 12.3 | 159.2 | 13.2 | - | - | 35.8 | 3.0 | 9.2 | 0.8 | 146.1 | 12.2 | 105.2 | 8.7 | 102.7 | 8.5 | - | - |
| 142.5 | 2.0 | 87.8 | 6.7 | 116.3 | 8.9 | 135.1 | 10.4 | - | - | 36.7 | 2.8 | - | - | 128.6 | 9.9 | 100.5 | 7.7 | 94.1 | 7.2 | - | - |
| 161.7 | 1.5 | 76.8 | 5.6 | 89.1 | 6.5 | 114.6 | 8.4 | - | - | 32.4 | 2.4 | - | - | 111.6 | 8.2 | 94.7 | 7.0 | 83.9 | 6.2 | - | - |
| 179.5 | 1.1 | 69.1 | 5.1 | 79.8 | 5.9 | 106.6 | 7.9 | - | - | 33.0 | 2.5 | - | - | 101.2 | 7.5 | 76.8 | 5.7 | 77.7 | 5.8 | - | - |
| 196.2 | 0.4 | 59.7 | 4.2 | 62.9 | 4.5 | 111.0 | 7.9 | - | - | 32.7 | 2.3 | - | - | 85.9 | 6.1 | 68.9 | 4.9 | 65.8 | 4.7 | - | - |

*Superscripts ($^c$, $^i$) refer to the type of cross section measured: c = cumulative, accounting for the decay of parent radionuclides to the quantified daughter; i = independent, considering direct formation only.

TABLE IV. Measured cross sections for the production of Gd radioisotopes by proton irradiation of terbium.

| Proton Energy (MeV) | ΔE (MeV) | Cross Section (mb)[*] | | | | | | | | | | | | | |
|---|---|---|---|---|---|---|---|---|---|---|---|---|---|---|---|
| | | $^{146}$Gd[c] | Δσ | $^{147}$Gd[c] | Δσ | $^{149}$Gd[c] | Δσ | $^{151}$Gd[c] | Δσ | $^{151}$Gd[i] | Δσ | $^{153}$Gd[c] | Δσ | $^{153}$Gd[i] | Δσ |
| 38.5 | 2.9 | - | - | - | - | - | - | - | - | - | - | 5.9 | 0.5 | - | - |
| 49.9 | 1.3 | - | - | - | - | - | - | - | - | - | - | 13.7 | 1.2 | - | - |
| 59.6 | 1.1 | - | - | - | - | - | - | 10.9 | 1.0 | - | - | 26.8 | 2.4 | - | - |
| 68.3 | 1.5 | - | - | - | - | - | - | - | - | - | - | 190.5 | 15.4 | - | - |
| 78.0 | 0.7 | - | - | - | - | 5.6 | 0.4 | 15.7 | 3.9 | 14.5 | 3.9 | 276.5 | 22.3 | - | - |
| 87.2 | 0.4 | - | - | 1.4 | 0.1 | 21.8 | 1.5 | 41.1 | 4.2 | 16.0 | 1.2 | 229.9 | 16.3 | - | - |
| 97.0 | 2.9 | - | - | 3.7 | 0.3 | 25.4 | 1.9 | 77.1 | 5.7 | 18.7 | 1.4 | 213.6 | 15.9 | - | - |
| 121.3 | 2.3 | 21.6 | 1.8 | 19.2 | 1.6 | 47.6 | 4.0 | 126.8 | 10.5 | 28.6 | 2.4 | 174.3 | 14.5 | 14.6 | 1.2 |
| 142.5 | 2.0 | 25.7 | 2.0 | 22.6 | 1.7 | 84.0 | 6.5 | 118.4 | 9.1 | 31.8 | 2.4 | 159.3 | 12.2 | 22.8 | 1.7 |
| 161.7 | 1.5 | 26.6 | 2.0 | 37.2 | 2.7 | 85.4 | 6.3 | 109.2 | 8.0 | 33.3 | 2.4 | 140.7 | 10.3 | 25.6 | 1.9 |
| 179.5 | 1.1 | 31.0 | 2.3 | 47.2 | 3.5 | 87.1 | 6.5 | 102.8 | 7.6 | 34.6 | 2.6 | 131.4 | 9.8 | 24.5 | 1.8 |
| 196.2 | 0.4 | 41.5 | 3.0 | 52.4 | 3.7 | 83.6 | 5.9 | 95.7 | 6.8 | 36.8 | 2.6 | - | - | - | - |

[*] Superscripts ('c', 'i') refer to the type of cross section measured: c = cumulative, accounting for the decay of parent radionuclides to the quantified daughter; i = independent, considering direct formation only.

TABLE V. Measured cross sections for the production of Eu radioisotopes by proton irradiation of terbium.

| Proton Energy (MeV) | ΔE (MeV) | Cross Section (mb)* | | | | | | | | | | |
|---|---|---|---|---|---|---|---|---|---|---|---|---|
| | | $^{145}$Eu$^c$ | Δσ | $^{146}$Eu$^i$ | Δσ | $^{147}$Eu$^c$ | Δσ | $^{148}$Eu$^c$ | Δσ | $^{149}$Eu$^c$ | Δσ | $^{149}$Eu$^i$ | Δσ |
| 38.5 | 2.9 | - | - | - | - | - | - | - | - | - | - | - | - |
| 49.9 | 1.3 | - | - | - | - | - | - | - | - | - | - | - | - |
| 59.6 | 1.1 | - | - | - | - | - | - | - | - | - | - | - | - |
| 68.3 | 1.5 | - | - | - | - | - | - | - | - | 0.1 | 0.0 | - | - |
| 78.0 | 0.7 | - | - | - | - | - | - | - | - | 7.3 | 0.6 | - | - |
| 87.2 | 0.4 | - | - | - | - | - | - | - | - | 27.6 | 1.9 | - | - |
| 97.0 | 2.9 | - | - | - | - | 5.6 | 0.4 | 2.3 | 0.2 | 27.6 | 2.1 | 3.8 | 0.3 |
| 121.3 | 2.3 | 2.5 | 0.2 | 4.7 | 0.4 | 31.6 | 2.6 | 10.0 | 0.8 | 60.4 | 5.0 | 7.6 | 0.6 |
| 142.5 | 2.0 | 12.9 | 1.0 | 12.8 | 1.0 | 37.0 | 2.8 | 11.5 | 0.9 | 102.7 | 7.9 | 11.2 | 0.9 |
| 161.7 | 1.5 | 20.0 | 1.5 | 15.5 | 1.1 | 56.4 | 4.1 | 12.8 | 0.9 | 98.8 | 7.3 | 10.5 | 0.8 |
| 179.5 | 1.1 | 24.8 | 1.8 | 18.0 | 1.3 | 69.0 | 5.1 | 14.5 | 1.1 | 109.6 | 8.1 | 13.3 | 1.0 |
| 196.2 | 0.4 | 25.2 | 1.8 | 20.1 | 1.4 | 74.3 | 5.3 | 15.9 | 1.1 | 109.1 | 7.8 | 17.0 | 1.2 |

*Superscripts ($^c$, $^i$) refer to the type of cross section measured: c = cumulative, accounting for the decay of parent radionuclides to the quantified daughter; i = independent, considering direct formation only.